\documentclass[aps,11pt,prc,preprint,superscriptaddress,nofootinbib]{revtex4}

\usepackage{CJK}
\usepackage[usenames]{color}
\usepackage{graphicx}
\usepackage{amsmath}
\usepackage{amsfonts}
\usepackage{amssymb}
\usepackage{mathrsfs}
\usepackage{bm}
\usepackage{verbatim}
\usepackage{cancel}

\newcommand{\beq}{\begin{equation}}
\newcommand{\eeq}{\end{equation}}
\newcommand{\bea}{\begin{eqnarray}}
\newcommand{\eea}{\end{eqnarray}}

\newcommand{\grad}{\vec{\nabla}}

\begin{document}

\title{Cutoff regulators in chiral nuclear effective field theory}
\author{Bingwei Long}
\email{bingwei@scu.edu.cn}

\author{Ying Mei}
\affiliation{Center for Theoretical Physics, Department of Physics, Sichuan University, 29 Wang-Jiang Road, Chengdu, Sichuan 610064, China}

\preprint{CTP-SCU/2016004}

\date{\today}

\begin{abstract}
Three-dimensional cutoff regulators are frequently employed in multi-nucleon calculations, but they violate chiral symmetry and Lorentz invariance. A cutoff regularization scheme is proposed to compensate systematically at subleading orders for these symmetry violations caused by regulator artifacts. This is especially helpful when a soft momentum cutoff has to be used for technical reasons. It is also shown that dimensional regularization can still be used for some Feynman (sub)diagrams while cutoff regulators are used for the rest.
\end{abstract}

\maketitle

\section{Introduction}

In applications of the chiral Lagrangian to multi-nucleon systems, ultraviolet (UV) three-momentum cutoffs, denoted by $\Lambda$, are often used in numerical calculations to regularize loop integrations over multi-nucleon intermediate states. Smooth or sharp, truncation of nucleon momenta can be thought of as momentum-dependent interactions, and they can be realized at the Lagrangian level by many insertions of operators involving derivatives acting on the nucleon fields. In order to preserve chiral symmetry, these ordinary derivatives must be part of chiral covariant derivatives [$SU(2)_L \times SU(2)_R$]:
\begin{equation}
    \mathscr{D}_\mu N \equiv \left(\partial_\mu + \frac{\bm{\tau}}{2} \bm{\cdot} \bm{E}_\mu\right) N \, , \label{eqn_covDN}
\end{equation}
where the so-called chiral connection operator is defined as
\begin{equation}
    \bm{E}_\mu \equiv i \frac{\bm{\pi}}{f_\pi}\bm{\times D}_\mu \, ,\label{eqn_connection}
\end{equation}
with
\begin{equation}
    \bm{D}_\mu \equiv D^{-1} \frac{\partial_\mu \bm{\pi}}{2f_\pi} \, ,\quad D \equiv 1 + \frac{\bm{\pi}^2}{4f_\pi^2} \, .
\end{equation}
Here $f_\pi \simeq 92$ MeV is the pion decay constant. Because the cutoff normally truncates three-momenta, Lorentz invariance is also broken by cutoff regulators. Therefore, it is a good strategy to deal with both symmetries at the same time.

Chiral symmetry and Lorentz invariance are expected to be better preserved with larger $\Lambda$'s (also to be shown more explicitly here), but that would imply more expensive calculations for multi-nucleon systems. Even for smaller systems where higher cutoffs can be afforded, in an era when nuclear physicists are pushing for high-precision calculations with chiral nuclear forces, we may want to think carefully about how to estimate and how to control these artificial symmetry breaking effects.

To various extent, this issue in the context of nucleon structure or nuclear forces has also been discussed in Refs~\cite{Bernard_2003rp, Epelbaum_2002gb}. Having been evaluated with cutoff regularization, the chiral expansion of certain quantities, namely, the nucleon mass and nucleon-nucleon scattering observables, is checked against their expected behavior based on chiral symmetry. Therefore, the inspection of symmetry-violating artifacts performed in Refs~\cite{Bernard_2003rp, Epelbaum_2002gb} relies on knowing in advance how chiral symmetry constrains the observables. The approach offered in the present paper starts with an effective Lagrangian that has symmetries \emph{and} regularization built in. Because symmetries and regularization are constructed simultaneously at the Lagrangian level, it is much less obligatory to check after calculations the observable against the symmetry constraints.

The manuscript is structured as follows. Section~\ref{sec_formal} introduces a specific cutoff regulator. In Sec.~\ref{sec_applications}, its applications to one- and two-nucleon systems demonstrates how to keep track of regulator-related symmetry violations. Finally, we summarize the main points in Sec.~\ref{sec_summary}.

\section{Framework\label{sec_formal}}

In order to recycle previous calculations done with dimensional regularization (DR), we will not fix the dimensionality to exactly four. A consistent way to have available both DR and a chiral-invariant cutoff regulator is to write formally the regulator into the Lagrangian with chiral covariant derivatives in $D$-dimension. The bare coupling constants will depend on the UV cutoff $\Lambda$ and $\mu$, which is the arbitrary mass scale appearing with DR: $\mathring{g}_i(\Lambda, \mu; \epsilon)$, where $2 \epsilon \equiv 4 - D$. One can for instance regularize the Lagrangian as follows:
\begin{equation}
S\left[\bm{\pi}, N^\dagger, N\right] = \int d^D x \left[\frac{1}{2}\left(N^\dagger \mathcal{O}^{(0)} e^{-\mathcal{O}^{(0)}/\lambda} N  + h.c. \right) + \cdots \right] \, , \label{eqn_lag}
\end{equation}
where
\begin{equation}
  \mathcal{O}^{(0)} \equiv i \mathscr{D}_0 + \frac{\vec{\mathscr{D}}^2}{2 m_N} + \cdots \, ,
\end{equation}
and the energy cutoff $\lambda$ is related to the momentum cutoff $\Lambda$ by
\begin{equation}
  \lambda \equiv \frac{\Lambda^2}{2m_N} \, .
\end{equation}
Here $\mathcal{O}^{(0)}$ is by construction chiral and Lorentz invariant, order by order in $m_N^{-1}$~\cite{Long_2010kt, Heinonen_2012km}.

Before we proceed to discuss this Lagrangian, it is worth digressing to a slightly academic note. Consider the path integral of Lagrangian~\eqref{eqn_lag}. The integration measure $[d\bm{\pi}][dN][dN^\dagger]$ is not necessarily chiral invariant, for $\bm{\pi}$ does not have to rotate as an isovector under chiral transformations in a nonlinear realization of chiral symmetry~\cite{Weinberg_1968de, Coleman_1969sm, Callan_1969sn}. This is not a bona fide anomaly of QCD Lagrangian, so a non-invariant term $\Delta \mathcal{L}$ must be added to cancel out the chiral violation of the integration measure. It is shown in the Appendix that $\Delta \mathcal{L}$ involves only the pion fields but not the nucleon fields, and it vanishes with DR. This is another motivation to hold on to DR, for pionic interactions can then be treated more easily.

To make use of this Lagrangian, we can expand $\mathcal{O}^{(0)}$ in powers of the pion fields:
\begin{equation}
\begin{split}
& \quad N^\dagger \left[i \left( \partial_0 + \frac{\bm{\tau}}{2} \bm{\cdot E}_0 \right) + \frac{\grad^2}{2m_N} + \cdots \right] e^{- \frac{\grad^2}{\Lambda^2}} \left[1 - \frac{i}{\lambda} \left( \partial_0 + \frac{\bm{\tau}}{2} \bm{\cdot E}_0 \right) + \cdots \right] N \\
= & N^\dagger \left(i\partial_0 + \frac{\grad^2}{2m_N} + \cdots \right) e^{- \frac{\grad^2}{\Lambda^2}} \left(1 - i\frac{\partial_0}{\lambda} + \cdots \right) N \\
& - \frac{1}{4f_\pi^2} N^\dagger \bm{\tau \cdot} \left(\bm{\pi \times} \dot{\bm{\pi}} \right) N + \frac{i}{4\lambda f_\pi^2} N^\dagger \bm{\tau \cdot} \left( \bm{\pi \times} \dot{\bm{\pi}}\right) \dot{N} + \cdots
\, ,
 \label{eqn_nonint}
\end{split}
\end{equation}
where the first term is purely kinetic and the rest contribute to pion-nucleon interactions. In constructing the kinetic term we have tactfully chosen to leave $-\grad^2/\Lambda^2$ (but not the time derivative) in the exponent of the Gaussian, whereas the Gaussian is expanded in $1/\Lambda$ everywhere else. The second term is the famed Weinberg-Tomozawa term.
The third one is the chiral connection accompanying the $\lambda$-suppressed time derivative, and it is the lowest-order operator to cancel chiral violations by the cutoff regulator. Although it has a nominal chiral index $\nu = 1$~\cite{Weinberg_1978kz}, it can often be relegated to $\nu = 2$ by invoking the equation of motion for nucleons, thanks to the time derivative acting on it. To justify the perturbative treatment of chiral connection operators, we recall one of the conclusions made in Ref.~\cite{Weinberg_1978kz}, that the pion loops are generally suppressed for external momenta much smaller than $4\pi f_\pi \simeq 1.2$ GeV.

The carefully arranged kinetic term translates into a nucleon propagator with the wanted suppression for nucleonic $3$-momenta:
\begin{equation}
  \frac{1}{i} S(p) = \frac{\exp \left( - \frac{\vec{p}\,^2}{\Lambda^2}\right)}{p_0 - \frac{\vec{p}\,^2}{2m_N} + \cdots + i\epsilon} \left(1 + \frac{p_0}{\lambda} + \cdots \right)
  \, . \label{eqn_Sp}
\end{equation}
Note that leaving $p_0/\lambda$ in the exponent would have caused any integral over $p_0$ to diverge. In the parentheses are operators to compensate for Lorentz violations of the Gaussian function.


\section{Applications\label{sec_applications}}

\subsection{One-nucleon processes\label{sec_1n}}

Consider the nucleon self-energy with incoming four-momentum $p$. The baryon energy is one order lower than the recoil correction, so the static-limit approximation can be used at leading order (LO). The loop integral of the sunset diagram for the self-energy is given by
\begin{equation}
  \int \frac{d^D l}{(2\pi)^D} \frac{\vec{l}\,^2}{l^2 - m_\pi^2} \frac{\exp \left[- (\vec{p}+\vec{l}\,)^2/\Lambda^2 \right]}{p_0 + l_0}\left[1 + \frac{p_0 + l_0}{\lambda} + \frac{(\vec{p}+\vec{l}\,)^2}{2m_N(p_0 + l_0)} + \cdots \right]  \, . \label{eqn_sunset_cutoff}
\end{equation}
The terms in the second bracket are to be computed order by order, with the counting rule: $p_0 + l_0 \sim |\vec{p} + \vec{l}\,| \sim Q$ and $\lambda \sim \Lambda_\chi^2/m_N$, where $Q$ refers generically to external momenta and $\Lambda_\chi \sim 1$ GeV is the chiral-symmetry-breaking scale.

If DR is to be used, the exponential factor will not be needed. So we expand the Gaussian in $1/\Lambda$, and return the integral to its more familiar form:
\begin{equation}
  \int \frac{d^D l}{(2\pi)^D} \frac{\vec{l}\,^2}{l^2 - m_\pi^2} \frac{1}{p_0 + l_0}\left[1 + \frac{p_0 + l_0}{\lambda} + \frac{(\vec{p}+\vec{l}\,)^2}{2m_N(p_0 + l_0)} - \frac{(\vec{p}+\vec{l}\,)^2}{\Lambda^2} + \cdots \right]  \, . \label{eqn_1nDR}
\end{equation}
Again, the terms in the bracket are to be computed perturbatively, and all of them need DR to be sensible. The $\Lambda$-suppressed terms, including $(p_0+l_0)/\lambda$ and $-(\vec{p}+\vec{l})^2/\Lambda^2$, would not have been there if only DR had been used from the very beginning. They provide the extra care one must give to subleading orders when both regulators are used. A technical note is in order regarding DR. It is convenient to write in evaluation that
\begin{equation}
   p_0 = v_\mu l^\mu \, , \quad \vec{p}\cdot\vec{q} = v\cdot p\, v\cdot q - p\cdot q \, ,
 \end{equation}
and to have $v = (1, \vec{0})$ when $D = 4$.

Integrals~\eqref{eqn_sunset_cutoff} and \eqref{eqn_1nDR} have different UV behaviors, so they require drastically different \emph{bare} nucleon mass $\mathring{m}_N$ to renormalize. With cutoff regularization~\eqref{eqn_sunset_cutoff}, $\mathring{m}_N$ has nontrivial running with respect to $\Lambda$, whereas it has a singularity like $1/\epsilon$ in renormalizing integral~\eqref{eqn_1nDR}. At any rate, after renormalization they will give similar results at each order, with discrepancy counted as higher-order effects.

We have gone to great lengths to have access to both DR and a cutoff regulator in a unified framework.
The most significant gain is that we can now reuse previous calculations using DR for some diagrams and in the mean time to use a cutoff regulator for others, e.g., in multi-nucleon processes.

\subsection{Two-nucleon processes\label{sec_2n}}

For definiteness, consider nucleon-nucleon scattering. The incoming (outgoing) momenta are denoted as $\vec{p}_1$ and $\vec{p}_2$ ($\vec{p}\,'_1$ and $\vec{p}\,'_2$). Define relative momenta and the center-of-mass energy as follows: $\vec{k} = (\vec{p}_1 - \vec{p}_2\,)/2$, $\vec{k}\,' = (\vec{p}\,'_1 - \vec{p}\,'_2\,)/2$, and $\overline{E} \equiv P_0 - \vec{P}\,^2/4m_N$, where $P_\mu \equiv (P_0, \vec{P})$ is the total four-momentum of the two-nucleon system.

The presence of pure nucleonic intermediate states, with four-momenta denoted by $(P_0/2 \pm l_0, \vec{P}/2 \pm \vec{l}\,)$, suggests that the nucleons can be very close to their mass shell. This invalidates even at LO the static-limit approximation for nucleons~\cite{Weinberg_1991um}. Stated in terms of power counting: $\vec{P}/2 \pm \vec{l} \sim Q$ and $P_0 \pm l_0 \sim Q^2/m_N$. This counting requires the recoil term $-\vec{p}\,^2/2m_N$ to be retained in the denominator of propagator~\eqref{eqn_Sp}, and resummation is necessary of the LO potential $V(\vec{k}\,', \vec{k}\,)$, which is the sum of one-pion exchange (OPE) and a few four-nucleon operators. The resummation in momentum space usually takes the form of the Lippmann-Schwinger equation.

If numerical calculations are inevitable in solving the Lippmann-Schwinger equation, it is necessary to regularize high three-momentum modes with the Gaussian in propagator~\eqref{eqn_Sp}.
Integrating out the zeroth component of the loop momentum gives the three-dimensional (off-shell) Lippmann-Schwinger equation:
\begin{equation}
  T^\lambda_{\overline{E}}\left(\vec{k}\,', \vec{k} \right) = V_\lambda\left(\vec{k}\,', \vec{k}\right) + \int \frac{d^3 l}{(2\pi)^3} V_\lambda\left(\vec{k}\,', \vec{l}\,\right) \frac{T^\lambda_{\overline{E}}\left(\vec{l}, \vec{k}\right)}{\overline{E} - \frac{l^2}{m_N} + i\epsilon} \, . \label{eqn_LSE}
\end{equation}
Here the regularized potential $V_\lambda(\vec{k}\,', \vec{k}\,)$ is related to $V(\vec{k}\,', \vec{k}\,)$, which is the sum of LO amputated, two-nucleon-irreducible diagrams, by
\begin{equation}
  V_\lambda\left(\vec{k}\,', \vec{k}\right) \equiv f_\lambda(k'; P) V\left(\vec{k}\,', \vec{k}\right) f_\lambda(k;P) \, ,
\end{equation}
where
\begin{equation}
  f_\lambda(k; P) = \exp \left(- \frac{\vec{P}\,^2}{4 \Lambda^2} \right) \exp \left(-\frac{\vec{k}\,^2}{\Lambda^2}\right) \, . \label{eqn_flambda}
\end{equation}
If there is no need to cut off contributions from large $\vec{P}$, one can even expand the exponent of Eq.~\eqref{eqn_flambda} in $\vec{P}^2/\Lambda^2$, which brings us to the more conventionally regularized Lippmann-Schwinger equation.

Besides the $\vec{p}\,^4/m_N^3$ correction from the denominator of the nucleon propagator~\eqref{eqn_Sp}, one needs to worry about the $p_0/\lambda$ term in the numerator. The contribution it gives to the next-to-next-to-leading potential has a simple structure:
\begin{equation}
  V_\lambda^{(2)}\left(\vec{k}\,', \vec{k}\right) = \frac{P_0}{\lambda} V_\lambda\left(\vec{k}\,', \vec{k}\right) + \cdots
\end{equation}

When calculating higher-order irreducible diagrams, we can use the same regularization scheme as in one-nucleon processes. One can follow the discussions in Sec.~\ref{sec_1n}, choosing either cutoff regulator \eqref{eqn_sunset_cutoff} or DR \eqref{eqn_1nDR} to regularize loop integrals.

\section{Discussions and Conclusions\label{sec_summary}}

Chiral symmetry and Lorentz invariance are often broken by cutoff regulators that truncate only three-momenta. Except for purely pionic systems, the symmetry-breaking artifacts of regulators can be mitigated by raising the momentum cutoff $\Lambda$. But in the cases where $\Lambda$ is somewhat soft, as in calculations for few- or many-nucleon systems, it is especially desirable to compensate for these symmetry violations order by order in $1/\Lambda$ and $1/m_N$.

The regularization scheme proposed here starts with Lagrangian~\eqref{eqn_lag}, with a ``kinetic'' term that is formally chiral and Lorentz invariant in $D$ dimensions. A careful arrangement was then made to obtain a nucleon propagator that suppresses high three-momentum contributions from the nucleons [see Eq.~\eqref{eqn_Sp}]. Expanding the chiral and Lorentz-invariant kinetic term generates a string of symmetry-breaking operators [see Eq.~\eqref{eqn_nonint}], which have fixed, $\Lambda$-dependent coefficients, and they will cancel order by order the symmetry breakings associated with the highly momentum-dependent propagator. Without them, we can no longer state at a given order that symmetry violations are higher-order effects.

Dimensional regularization can still be used. As demonstrated in Sec.~\ref{sec_1n}, once the choice of regulator is made at LO, the consequence of choosing a certain regulator will arise in subleading orders. A nontrivial benefit is that we can recycle previous DR-based calculations for, say, one-nucleon processes and irreducible pion-exchange diagrams. In the mean time, cutoff regulators are used for resummation in multi-nucleon processes. The price is the extra cares we must give to higher-order calculations, as demonstrated in Sec.~\ref{sec_applications}.

The justification of the method uses a Lagrangian path integral formalism; therefore, the unitarity of the $S$ matrix is not manifest at the beginning, as opposed to canonical quantization. But we can always verify the unitarity by checking explicitly whether the amplitudes obey the optical theorem. Such a check has been done to integral equation~\eqref{eqn_LSE}.

One may want to design other cutoff regularization schemes for various purposes.  The important points to keep in mind are that (a) the integration over three-momenta is regularized as desired, (b) we have a generating device to keep track of chiral and Lorentz violations in a systematic fashion, and (c) it does not interfere with counting Feynman diagrams in powers of external momenta.

Since the regularization scheme devised here is merely one possibility, its features may not be shared by other consistent schemes. For instance, it has only one cutoff value, but other schemes could be imagined in which multiple cutoff values are utilized. One can for example attach a chiral and Lorentz-invariant form factor to every Lagrangian interaction operator, each with a distinctive cutoff value. With such a construction, a possible consequence is that part of three-nucleon forces will be regularized differently than two-nucleon forces.

The regularization scheme presented here will lead to nonlocal pion-exchange nucleon-nucleon potentials. References~\cite{Epelbaum_2014efa, Epelbaum_2014sza} argue that local cutoff regulators are preferable for pion exchanges because local regulators do not distort the analytical structures of pion-exchange diagrams, although several quantum Monte Carlo calculations found strong regulator dependence in neutron matter when local regulators are used~\cite{Gezerlis_2013ipa, Gezerlis_2014zia}. It remains an open issue as to why analyticity and unitarity impose more constraints on regularization schemes than symmetries and power counting do. (For further reading on nucleon-nucleon scattering using the analytic properties and unitarity of the $S$ matrix, see Refs.~\cite{Gasparyan_2012km, Guo_2013rpa}.)

\textit{Note added in proof:} After the manuscript had been accepted, we noticed that Ref.~\cite{Djukanovic_2004px} had employed an idea similar to ours. But there are significant differences in the implementation. Heavy-baryon formalism was used here from the beginning, and order-by-order preservation of Lorentz invariance was carefully demonstrated. We chose a different regulator, the Gaussian function in three-momenta, as the example, which is more often used in practical calculations.

\acknowledgments

We thank Bira van Kolck for reminding us of the early works~\cite{Charap_1970xj, Salam_1971sp, Honerkamp_1971va, Gerstein_1971fm} on related issues and the Institut de Physique Nucl\'eaire d'Orsay for the hospitality when part of the work was carried out there. This work was supported in part by the National Natural Science Foundation of China (NSFC) under Grant No. 11375120.

\appendix*

\section{$\Delta \mathcal{L}$ is purely pionic\label{app_nobaranon}}

The path integral of chiral effective field theory is
\begin{equation}
\int [d\bm{\pi}][dN][dN^\dagger]\, \exp \left[ i\int d^D x\, \mathcal{L}\left(N^\dagger, N, \bm{\pi}\right) + \Delta \mathcal{L}(\bm{\pi}) \right] \, ,
\end{equation}
where the Lagrangian $\mathcal{L}\left(N^\dagger, N, \bm{\pi}\right)$ is invariant under nonlinearly realized chiral transformation~\cite{Weinberg_1968de}, parametrized by $\bm{\theta}_A$:
\begin{align}
    \delta_A \bm{\pi} &\equiv \bm{\pi}_\star - \bm{\pi} = f_\pi \bm{\theta}_A \left(1 - \frac{\bm{\pi}^2}{4f_\pi^2}\right) + \bm{\theta}_A \bm{\cdot} \frac{\bm{\pi}}{2f_\pi} \bm{\pi} \, , \label{eqn_pitrans}\\
    \delta_A N &\equiv N_\star - N = i \left(\bm{\theta}_A \bm{\times} \frac{\bm{\pi}}{2f_\pi} \right) \bm{\cdot} \frac{\bm{\tau}}{2} N \, . \label{eqn_Ntrans}
\end{align}
Here $N$ is a heavy-baryon field instead of a Dirac field, so it has a two-valued spin index instead of Dirac ones. Upon a chiral transformation, $N$ undergoes a spacetime-dependent isospin rotation, parameterized by $\bm{\theta}_A \bm{\times} \bm{\pi}/2f_\pi$; on the contrary, the chiral transformation of $\bm{\pi}$ does not resemble in any way a spacetime-dependent isospin rotation. Because of that, as it turns out, the measure is not chiral invariant, and it requires a chiral-non-invariant term $\Delta \mathcal{L}$ to neutralize the violations. We wish to show here that $\Delta \mathcal{L}$ depends only on the pion fields: $\Delta \mathcal{L}(\bm{\pi})$. References~\cite{Charap_1970xj, Salam_1971sp, Honerkamp_1971va, Gerstein_1971fm} discussed similar issues, however, without baryonic degrees of freedom.

We can choose to integrate over the baryonic degrees of freedom before dealing with the pion fields. Noticing that $\bm{\pi}_\star(x)$ depends on $\bm{\pi}(x)$, but not on $N(x)$,
\begin{equation}
    \bm{\pi}_\star(x) = \bm{\pi}_\star \left[\bm{\pi}(x); \bm{\theta}_A\right] \, ,
\end{equation}
one can transform first $[dN][dN^\dagger]$ to $[dN_\star][dN_\star^\dagger]$ with $\bm{\pi}(x)$, or $\bm{\pi}_\star(x)$, as fixed parameters. Therefore, the Jacobian of the whole measure factorizes into two parts:
\begin{equation}
    [d\bm{\pi}_\star][dN_\star][dN_\star^\dagger] = J_\pi(\bm{\pi}; \bm{\theta}_A) J_N\left(N, N^\dagger; \bm{\pi}, \bm{\theta}_A\right) [d\bm{\pi}][dN][dN^\dagger] \, ,
\end{equation}
where $J_N$ is defined as if $\bm{\pi}(x)$ were unchanged in the transformation~\eqref{eqn_Ntrans}:
\begin{align}
    [dN_\star][dN_\star^\dagger] = J_N\left(N, N^\dagger; \bm{\pi}, \bm{\theta}_A\right) [dN][dN^\dagger] \, .
\end{align}
$J_\pi$ is defined according to the transformation~\eqref{eqn_pitrans} regardless of presence of the baryon fields:
\begin{equation}
  \begin{split}
    J_\pi \left(\bm{\pi}; \bm{\theta}_A\right) &= \det \left(\frac{\partial \pi_\star^a}{\partial \pi_b}\, \delta^{(D)}(x - x')  \right)  \, , \\
    &= \exp \left[ \frac{N_f}{2f_\pi} \int d^D x\, \delta^{(D)}(0)\, \bm{\theta}_A \bm{\cdot \pi}  \right] \, ,
  \end{split}
\end{equation}
where $N_f = 3$ is the number of flavor of the pions.

The fermionic Jacobian $J_N$ is related to the transformation matrices as follows:
\begin{equation}
    J_N = (\det U\, \det U^c)^{-1} \, ,
\end{equation}
where
\begin{align}
    U_{a\sigma x,\, a'\sigma' x'} &\equiv \left[1 - i \left(\bm{\theta}_A \bm{\times} \frac{\bm{\pi}}{2f_\pi} \right) \bm{\cdot} \frac{\bm{\tau}}{2} \right]_{a\, a'} \delta_{\sigma \sigma'}\, \delta^{(D)}(x - x') \, , \\
    U^c_{a\sigma x,\, a'\sigma' x'} &\equiv \left[1 + i \left(\bm{\theta}_A \bm{\times} \frac{\bm{\pi}}{2f_\pi} \right) \bm{\cdot} \frac{\bm{\tau}}{2} \right]_{a\, a'} \delta_{\sigma \sigma'}\, \delta^{(D)}(x - x') \, ,
\end{align}
with $a (a')$ being isospin indecies, $\sigma (\sigma')$ spin indecies, and $x (x')$ spacetime coordinates. It is easy to verify that $U^c$ is the hermitian conjugate of $U$,
\begin{equation}
    U^c = U^\dagger \, ,
\end{equation}
and that $U$ is unitary,
\begin{equation}
    \left(U U^c \right)_{a\sigma x,\, a'\sigma' x'} = \delta_{a\, a'} \delta_{\sigma \sigma'}\, \delta^{(D)}(x - x') \, .
\end{equation}
It follows immediately that $J_N$ is unity:
\begin{equation}
    J_N = 1 \, .
\end{equation}
Therefore, the overall integral measure transforms as if the nucleon fields were not present, and $\Delta \mathcal{L}$ needs to cancel out only $J_\pi(\bm{\pi}; \bm{\theta}_A)$. Therefore, it must have the following form:
\begin{equation}
  \Delta \mathcal{L} = i \delta^{(D)}(0)\, g(\bm{\pi}^2/4f_\pi^2) \, ,
\end{equation}
with $g(x)$ satisfying
\begin{equation}
  \frac{d g}{d x} = N_f \left(1 + x\right)^{-1} \, .
\end{equation}
Solving for $g(x)$, we arrive at
\begin{equation}
  \Delta \mathcal{L}(\bm{\pi}) = i \delta^{(D)}(0) N_f \ln \left(1 + \frac{\bm{\pi}^2}{4 f_\pi^2} \right) \, .
\end{equation}
As promised, $\Delta \mathcal{L}$ is found to be independent of the baryon fields and it vanishes with DR.

Although we have used a particular parametrization for nonlinear realization of chiral symmetry, the above conclusion will hold for other parametrizations, provided that the pion fields transform without any reference to the baryon fields and chiral transformation of the baryon fields is realized as an unbroken rotation with angles depending only on the local values of the pion fields. Any parametrization schemes following the renowned Callan-Coleman-Wess-Zumino (CCWZ) technique~\cite{Coleman_1969sm, Callan_1969sn} will meet the above requirements.

\end{document}